\documentclass[aps,prb,twocolumn,superscriptaddress]{revtex4-2}
\usepackage[colorlinks=true, urlcolor=blue, linkcolor=blue, citecolor=blue, pdftex]{hyperref}
\usepackage{xr} 
\usepackage[dvipsnames]{xcolor}
\usepackage[utf8]{inputenc}
\usepackage[normalem]{ulem}
\usepackage{braket}
\usepackage{graphicx}
\usepackage{caption}
\usepackage{subcaption}

\usepackage{amsmath}
\usepackage{amsfonts} 
\usepackage[capitalise]{cleveref}

\begin{document}

\title{On the Optimal Layout of Two-Dimensional Lattices for Density Matrix Renormalization Group}

\date{\today}

\begin{abstract}
For quantum spin models defined on a two-dimensional lattice, we look for the best numbering of the lattice sites (a layout) that, at fixed bond dimension and other parameters of the density matrix renormalization group (DMRG) algorithm, gives the lowest value of the variational energy, maximum entropy and truncation error. We consider the conjecture that the optimal layout is a Hamiltonian path, and that it optimizes a simply computable geometric cost function. Finding the minimum of such a function, which is a variant of the minimum linear arrangement problem, provides the DMRG with an efficient layout of the lattice and improves both accuracy and convergence time. Applications to antiferromagnetic or spin glass spin-$1/2$ models on the square and triangular lattices are studied.
\end{abstract}

\author{Antonello Scardicchio}
\affiliation{The Abdus Salam ICTP, Strada Costiera 11, 34151 Trieste, Italy}
\affiliation{INFN, Sezione di Trieste, Via Valerio 2, 34127 Trieste, Italy}

\maketitle
\section{Introduction}

Two-dimensional quantum lattice models --such as the Hubbard model, the Heisenberg antiferromagnet, various Ising or Heisenberg spin glasses, and lattice gauge theories-- occupy a critical position in condensed matter physics \cite{manousakis1991spin,lee2006doping,balducci2022localization,viteritti2025quantum,sachdev2025lectures}. Two dimensions is both experimentally accessible and a theoretical sweet spot: Much interesting {\it quantum} physics happens in two dimensions, far from the mean-field predictions but at the same time not too far from one dimension, so that modern numerical methods which are best suited for one dimension, like density matrix renormalization group (DMRG) \cite{white1992density,schollwock2005density}, matrix product states or operators (MPS-MPO)\cite{verstraete2023density}, projected entangled pairs states (PEPS) \cite{verstraete2008matrix} can still be applied with a certain degree of success.

\begin{figure}
    \centering
    \includegraphics[width=0.7\linewidth]{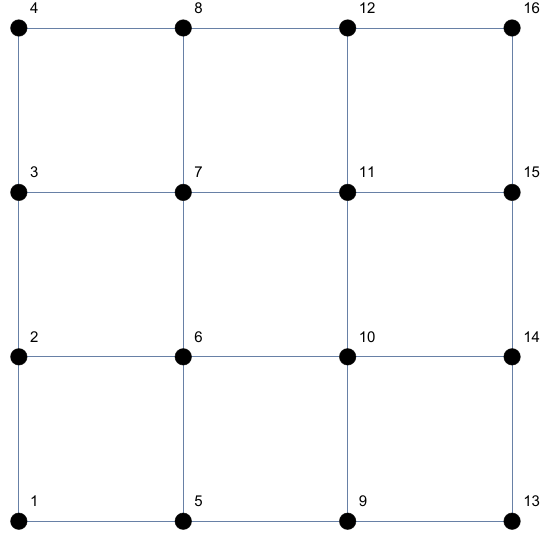}
    \includegraphics[width=0.9\linewidth]{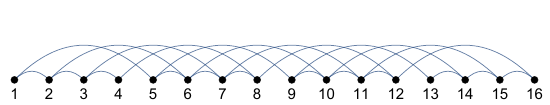}
    \caption{A square lattice with a simple layout. The layout is the numbering of the vertices, which induces the graph on the bottom.}
    \label{fig:square_ex}
\end{figure}

DMRG and Tensor Network \cite{orus2019tensor,felser2020two,gerster2017fractional} methods are, in fact, naturally formulated for one-dimensional chains but can be adapted with some degree of success to higher-dimensional models \cite{stoudenmire2012studying}, by mapping them to a one-dimensional chain with long-range interactions. Mapping a higher-dimensional lattice to a one-dimensional one is an instance of {\it graph layout} \cite{diaz2002survey}. A graph layout trivially exists for any graph, see Fig.\ref{fig:square_ex}, however deciding whether the layout has a particular property or finding the layout that, for a given graph, is optimum with respect to some measure, is a hard combinatorial problem, often in the NP-hard or NP-complete class. Graph layout problems have received a lot of attention in very diverse fields. In computer science, the problem appears in circuit design and computer network optimization, in bioinformatics, graph layout algorithms are used to project high-dimensional data, such as protein-protein interaction networks and single-cell gene expression profiles, into interpretable visualizations that reveal functional clusters and cellular relationships \cite{wolf2018scanpy}.

In this sense, the problem we consider in this paper follows a long tradition of incarnations of the graph layout problem. We want to investigate the graph layout for mapping two-dimensional lattices into one-dimensional spin chains, and in particular we want to optimize for the performance of the algorithm. Recently, there have been several heuristic solutions for an optimal layout for the DMRG algorithm \cite{cataldi2021hilbert,magnifico2025tensor,abedi2025efficient,bellwood2025fractal,pavevsic2025scattering} which showed promise to improve both the ground state detection and the time evolution of initial states for quantum simulations. In those papers, there is already the idea that the optimal layout should be a fractal path, which wanders in a local patch of sites, before leaping to a neighboring patch and doing that again. Here we start from this generic consideration and go further in various directions trying to answer the final question {\it what is the optimal layout for running the DMRG algorithm?}

One should first consider two fundamental questions: {\it 1) Is the optimal graph layout for DMRG a Hamiltonian path?} We have not investigated this question systematically but from a few instances, it seems that the short answer is yes. This is not obvious {\it a priori} since the optimal layout depends strongly on the cost function we need to optimize. For example, it is known for some natural cost functions, that the answer is {\it negative} \cite{mitchison1986optimal}. The optimal layout is not a path, and big jumps are necessary to optimize the cost function. However, for the best performance of the DMRG it seems one can restrict to Hamiltonian paths instead of general layout. The investigation of this problem has been hampered by the high cost of running DMRG for every proposed layout.

This brings us to a second problem: {\it 2) Is there a geometric cost function which is so strongly correlated with the performance of DMRG that we can use it as a proxy of the latter?} This is necessary, as we said, for a comprehensive study of the optimal layout/path since in an optimization algorithm, evaluating the cost function is something which is done at every proposed change in the proposal. Instead, if we find a geometric cost function that can be evaluated efficiently for the given layout and only call DMRG on the optimal path, one can get a significant speedup.

Let us introduce some fundamental concepts in the next section and then dwell in the optimization problem in the following one. 

\section{Layouts and Optimal Paths}

\subsection{General considerations}

We follow \cite{diaz2002survey} defining the graph $G(V,E)$ where $V=\{1,...,N\}\subset\mathbb{N}$ is the set of vertices, and $E=\{uv| u,v\in V\}$ is the collection of the edges. A layout is a bijection of the vertices $\varphi:V\ni i\mapsto j\in\{1,...,N\}.$ A Hamiltonian path is a particular layout in which any two consecutive vertices $j,j+1$ are joined on the original graph $\varphi^{-1}(j)=u,\varphi^{-1}(j+1)=v,\ uv\in E$. Given a layout $\varphi$ there are a plethora of cost functions one can define, the optimization of which can lead to complicated problems. We are interested in one of them. 

Define the layout distance between two vertices $u,v\in V$, $\lambda(u,v)=|\varphi(u)-\varphi(v)|$. In principle, this distance can go up to $N-1$. If $uv\in E$ (the two vertices are connected on $G$), then $\lambda(u,v)$ is the called the {\it length} of $uv$ in the layout $\varphi$.

The linear arrangement $\mathrm{LA}$ is defined as
\begin{equation}
    \mathrm{LA}(\varphi,G)=\sum_{uv\in E}\lambda(u,v).
\end{equation}
This leads to the problem $\textsc{MinLA}(G)=\min_{\varphi}\textsc{LA}(\varphi,G)$ originally introduced in \cite{harper1964optimal} and solved for hypercubes in \cite{mitchison1986optimal}: the optimal layout is {\it not} a Hamiltonian path. 

One can generalize the cost function as follows
\begin{equation}
    \mathrm{LA}_q(\varphi,G)=\sum_{uv\in E}\lambda(u,v)^q,
    \label{eq:LAq}
\end{equation}
with $q<1$ and the corresponding $\textsc{MinLA}_{q}(G)$. This has been already considered in \cite{mitchison1986optimal} where several lower bounds are given for the optimal layout. We will see that the cost function which is best correlated with the performance of the DMRG algorithm is $\textsc{LA}_{1/2}$, which, as already recongized in \cite{mitchison1986optimal} is a particular case, separating the $1\geq q>1/2$ and the $q<1/2$ cases which have a simpler behavior of the optimal layouts.

\subsection{Cost function for DMRG}

The layout problem in DMRG appears for the following reason \cite{white1992density}. Consider a system of $N$ spin-$1/2$ degrees of freedom, whose Hamiltonian has interactions which define some particular graph $G$. The MPS ansatz with a given bond dimension $\chi$ for a given wave function $\ket{\Psi}$ is
\begin{equation}
    \ket{\Psi}=\sum_{\{s\}}A^{1}_{s_1}\cdot A^{2}_{s_2}\cdot...\cdot A^{N}_{s_N}\ket{s_1,...,s_N}
\end{equation}
where each $A^i_{s_i}$ is a  $\chi\times\chi$ matrix indexed by the spin-$z$ value $s_1=\pm 1/2$. This is independent of the dimension of the lattice on which the function is defined, and the numbering of the spins is a layout of the graph corresponding to the lattice of the interactions. This makes the MPS ansatz natural for one-dimensional systems, where $G$ is a line, on which it gives unparalleled accuracy for the ground state. In order to solve for generic $G$, for example two- and higher- dimensional systems, one needs a layout $\varphi(G)$. 

\begin{figure}
    \centering
    \includegraphics[width=0.45\linewidth]{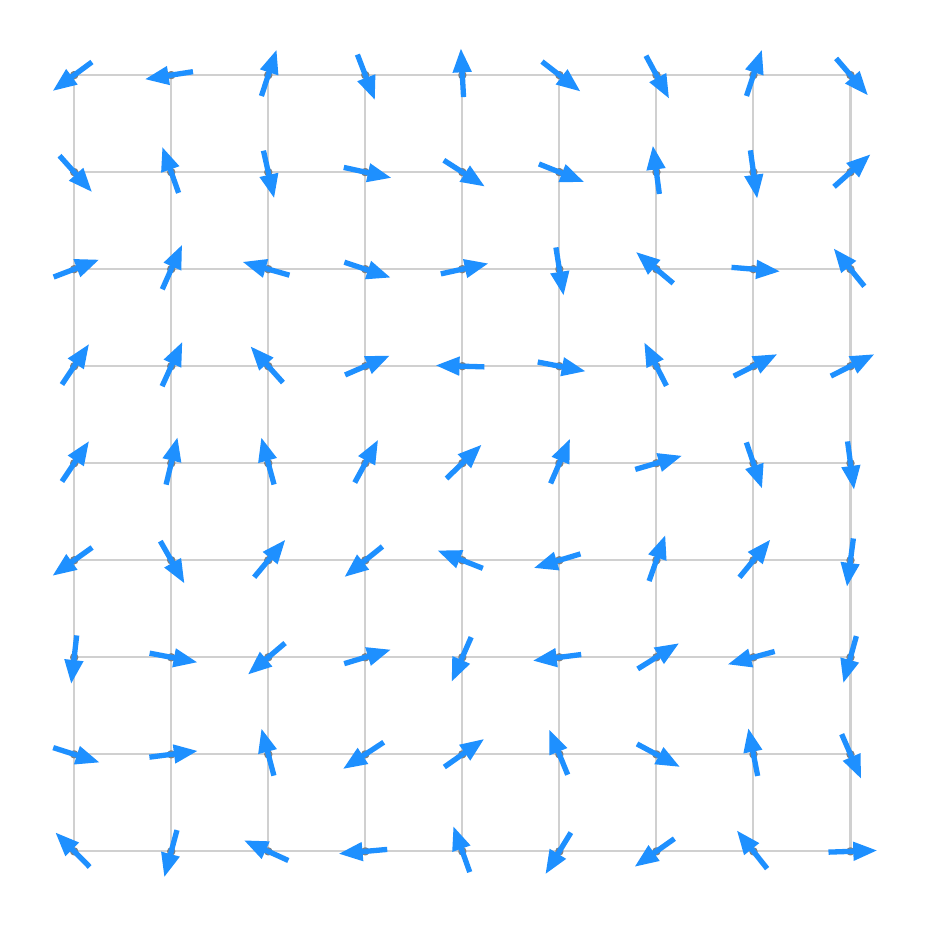}
    \includegraphics[width=0.45\linewidth]{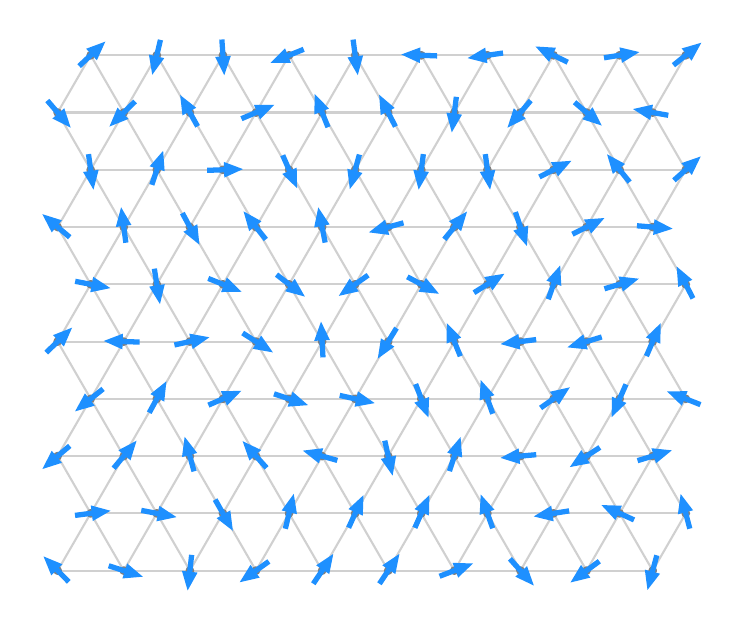}
    \caption{Square lattice for the definition of the antiferromagnetic model in Eq.s(\ref{eq:SqLAFM}) and (\ref{eq:SqL_SG}), and triangular lattice for the model in Eq.(\ref{eq:TlAFM}). The little arrows represent the spin degrees of freedom.}
    \label{fig:Square_Triang}
\end{figure}

To be concrete we will consider two problems (see Fig.\ref{fig:Square_Triang}) on the square lattice with open boundary conditions: $G\sim\mathbb{Z}_L\otimes\mathbb{Z}_L$. The first is the square lattice, spin-$1/2$ antiferromagnet:
\begin{equation}
    H=J\sum_{\langle uv\rangle}\vec{s}_u\cdot\vec{s}_v,
    \label{eq:SqLAFM}
\end{equation}
where $\langle uv\rangle$ is the traditional notation for the edges $uv\in E$, of the square lattice and the second is the square lattice Heisenberg spin glass studied in \cite{viteritti2025quantum} using the method of \cite{rende2025foundation}:
\begin{equation}
    H=\sum_{\langle uv\rangle}J_{uv}\vec{s}_u\cdot\vec{s}_v,
    \label{eq:SqL_SG}
\end{equation}
where $J_{uv}=\pm 1$ with probability $1/2$. Both problems are notoriously resilient to numerical investigation, the second one suffering of a severe sign problem as well.

To change the lattice we also consider the antiferromagnet on a triangular lattice, and allow for anisotropies in the couplings:
\begin{equation}
   H=J_1\sum_{\langle uv\rangle} \vec{s}_u\cdot\vec{s}_v+J_2\sum_{\langle\langle uv\rangle\rangle} \vec{s}_u\cdot\vec{s}_v
\label{eq:TlAFM}
\end{equation}
where $\langle uv\rangle\in E$ are the edges that would reconstruct the square lattice, and $\langle\langle uv\rangle\rangle\in E$ are the edges which would correspond to a diagonal on the square lattice (the traditional $J_1-J_2$ model has $J_2$ on all the nearest neighbors couplings). The $J_2$ coupling softens the antiferromagnetic ordering on the square lattice produced by $J_1$.\footnote{Notice this is a variant of the $J_1-J_2$ AFM model since the $J_2$ couplings do not extend to next-nearest neighbors.}

Let us consider the first case, the antiferromagnet on the square lattice Eq.(\ref{eq:SqLAFM}). The traditional solution for the choice of the mapping has been to use a simple and safe {\it snake path}, namely choosing $u=(0,0)$ and $v=(L-1,L-1)$ (or $(0,L-1)$ depending on the parity of $L$) that scans the lattice row by row. While such a path has the advantage of adding a negligible overhead to the computation, it has the disadvantage that it is almost certainly {\it not} the optimal path. When possible, however, one can compensate by increasing the bond dimension, the number of sweeps or introducing other kinds of optimizations in the algorithm. The inefficiency of choosing the snake path becomes a burden if one wants to go to larger $L$. It is therefore not a surprise that in the recent literature, there has been the observation that some curves (a generalized Hilbert curve\cite{abedi2025efficient} or other kind of curves \cite{bellwood2025fractal}), can give a computational advantage with equally modest computational overhead. 

\begin{figure}
    \includegraphics[width=0.95\linewidth]{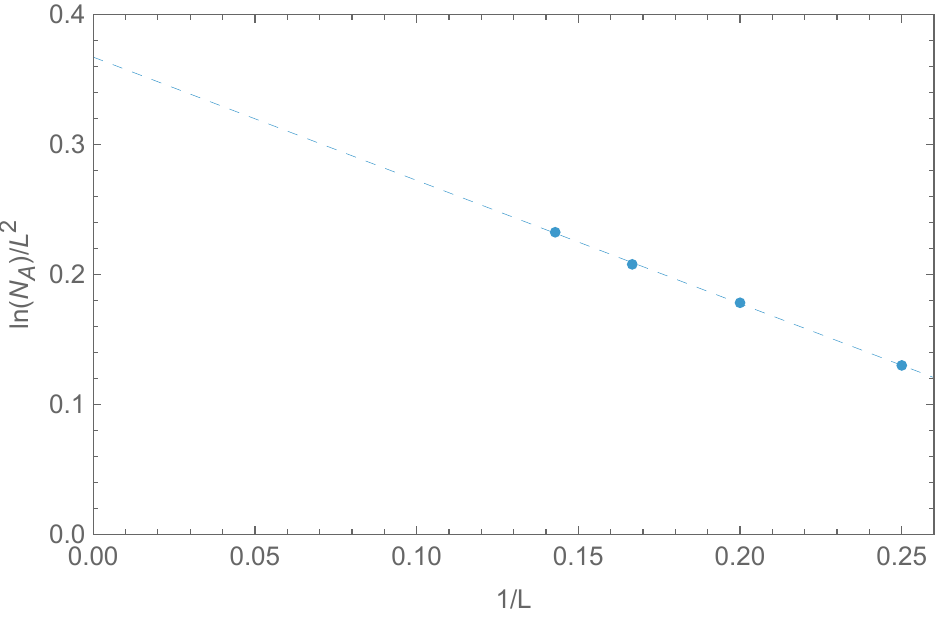}
    \caption{The number of paths obtained with exact enumeration with fixed extrema on adjacent sides $L=4,...,7$. The exponential fit $\ln(N)/L^2=0.36-0.95/L$. The leading term agrees within error with that for the total number of paths $0.352$ in \cite{jacobsen2007exact}.}
    \label{fig:entropyHam}
\end{figure}

Hamiltonian paths have been studied on regular lattices as models of compact polymers or self-avoiding walks. Efficient algorithms exist to sample them which lead to their exact enumeration for small lattices \cite{jacobsen2007exact,oberdorf2006secondary}, and a statistical mechanics, partition function calculation of their proliferation can be set-up and solved in the limit of large connectivity \cite{orland1985evaluation}. The result of these analyses is that the number of Hamiltonian paths scales like
\begin{equation}
    \mathcal{N}\sim e^{s L^2},
\end{equation}
with the entropy density $s\simeq 0.352$ (to be compared with the large-connectivity prediction \cite{orland1985evaluation} $s_{\mathrm{MF}}=\ln(4)-1=0.38629...$). Of these paths, however, we will need only those with starting and ending point on the boundary. Ideally, on the corners. The reason for this will be clear in the following. This reduces the number of paths considerably, however the exact number is not known (it should not be difficult to calculate, though), but a quick numerical calculation of {\it entropy density} $s=\lim_{L\to\infty}\ln\mathcal{N}/L^2$ gives a value compatible with the numbers above (see Figure \ref{fig:entropyHam}).

The ideal cost function would be $C_{\mathrm{DMRG}}[\varphi,\chi]=E_0(\chi)$ the energy of the ground state for a given layout, at fixed bond dimensions $\chi$. However, computing $E_0$ is computationally vexing, in particular for $\chi\gtrsim 150$ or so and $L\gtrsim 10$. So one must find a different cost function which is correlated with $C_{\mathrm{DMRG}}$ and that can be easily computed. If we succeed in doing so, we find the optimum path $\varphi^*$ and use that as an input for the DMRG algorithm, pushing the calculations to large $\chi$. The gain in energy $E_0(\chi)$ can be consistent, if the path is chosen appropriately. From now on we will drop the dependency of $\varphi$ on the initial and final points, as we will fix them to be distance $L-1$ apart on the lattice.

\begin{figure*}[t]
    \includegraphics[width=0.32\linewidth]{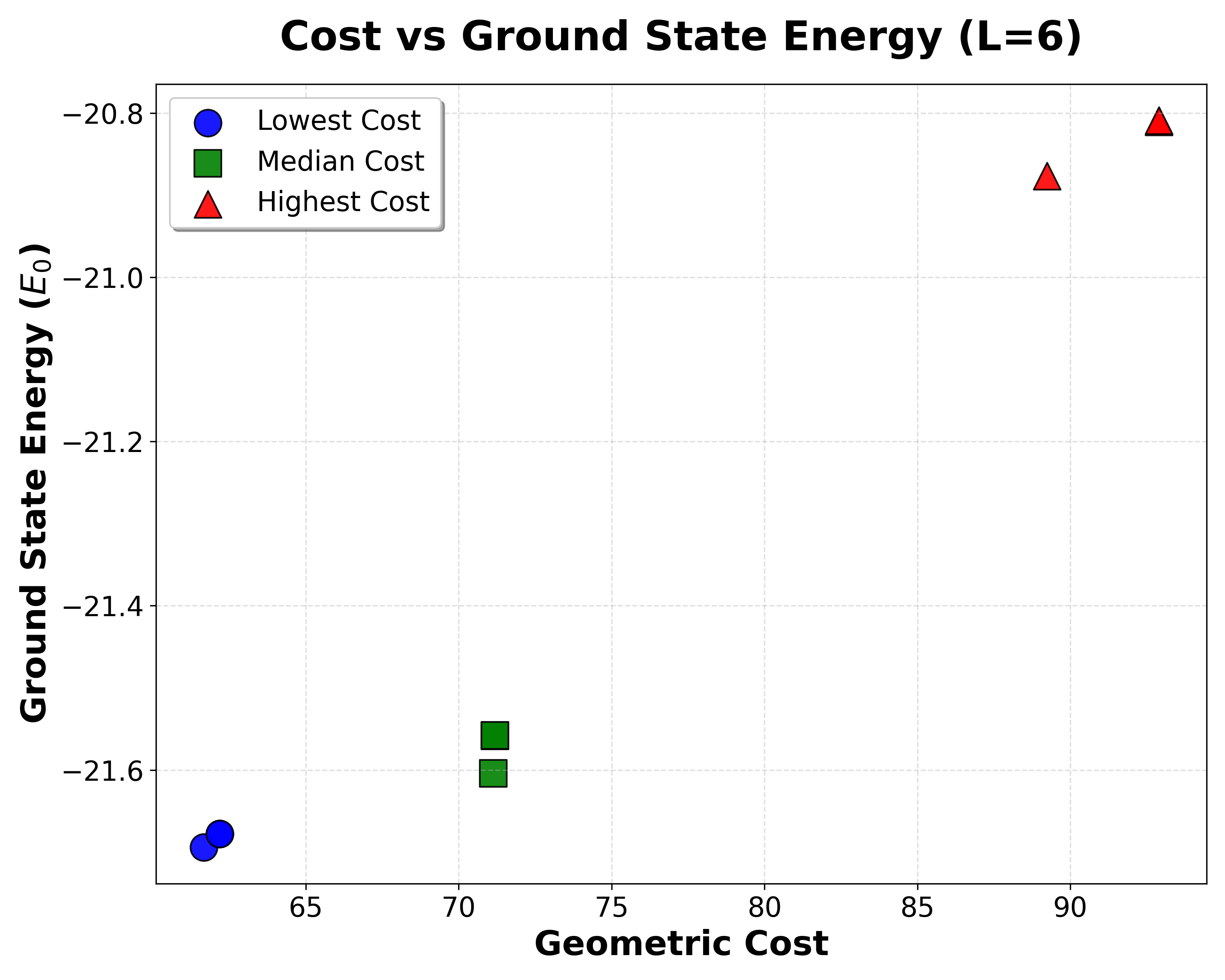}
    \includegraphics[width=0.32\linewidth]{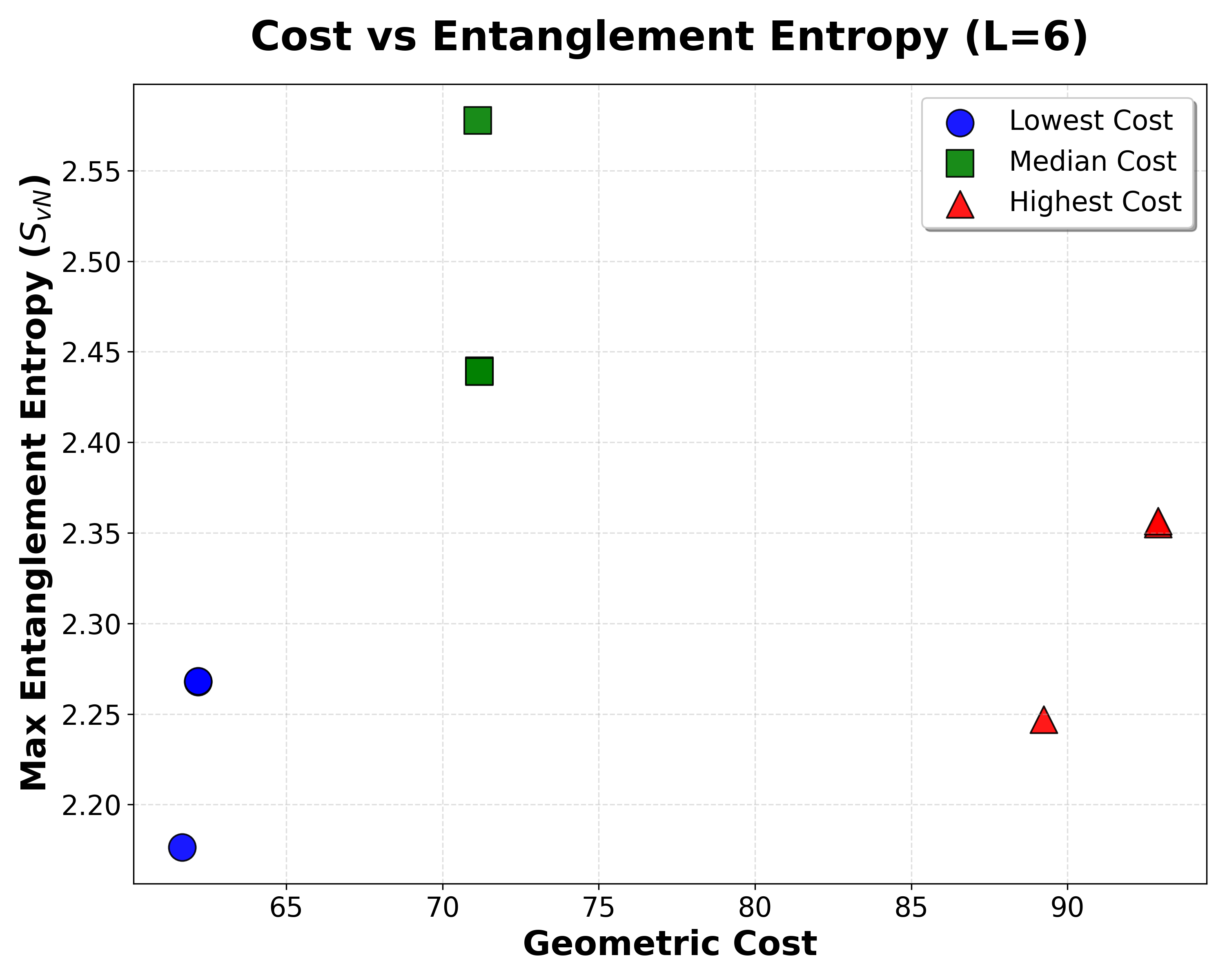}
    \includegraphics[width=0.32\linewidth]{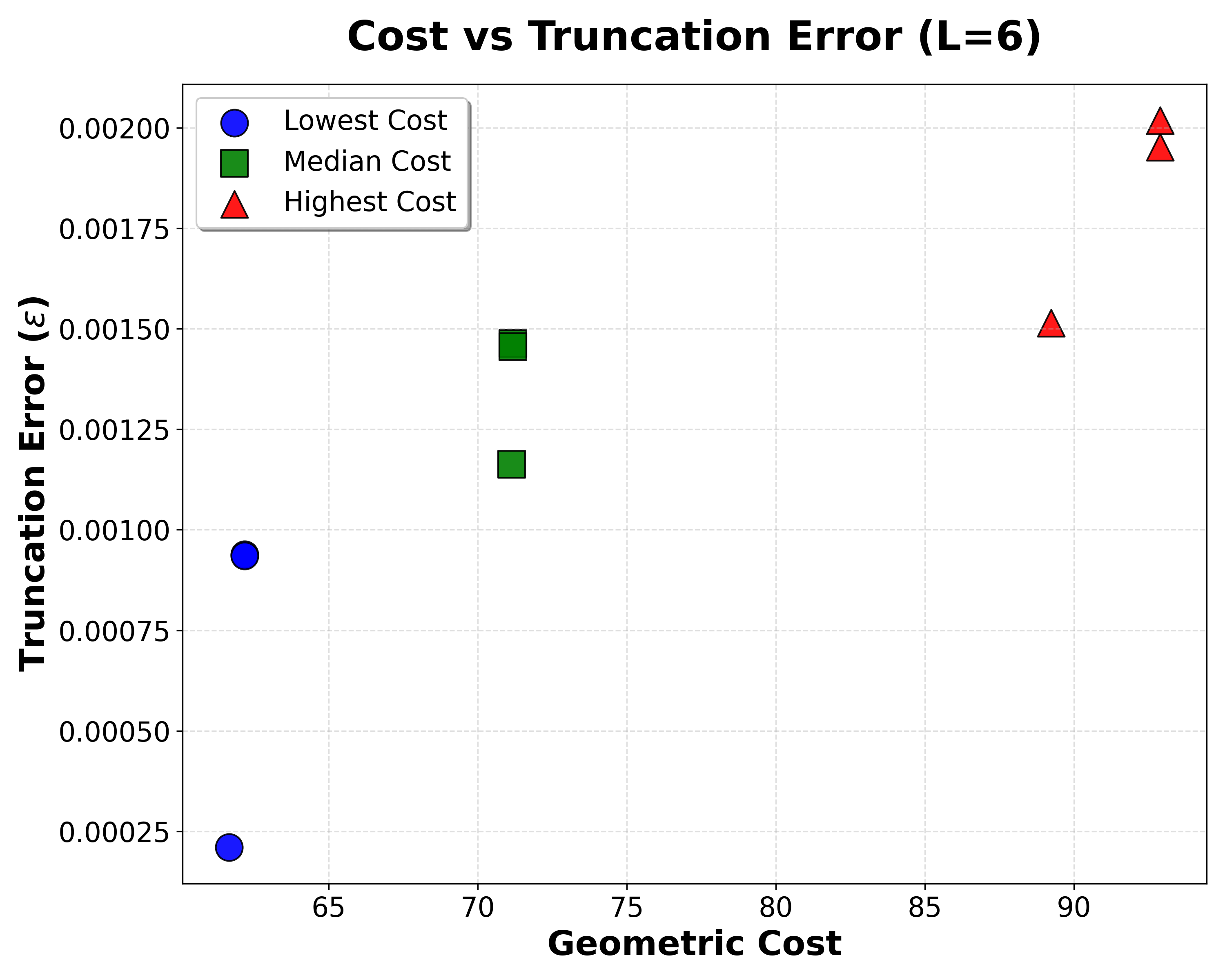}
    \caption{Square lattice antiferromagnet: Correlation between the geometric lowest cost in Eq.(\ref{eq:CLA12}) and its best performance of the DMRG algorithm for $L=6$. All the paths in $L=6$ are found but only 9 of them are selected (some points are coincident). DRMG uses TenPy with Maximum bond dimension $\chi=100$, maximum number of sweeps 100.}
    \label{fig:Correlation_C_DMRG}
\end{figure*}

\begin{figure}
    \includegraphics[width=0.95\linewidth]{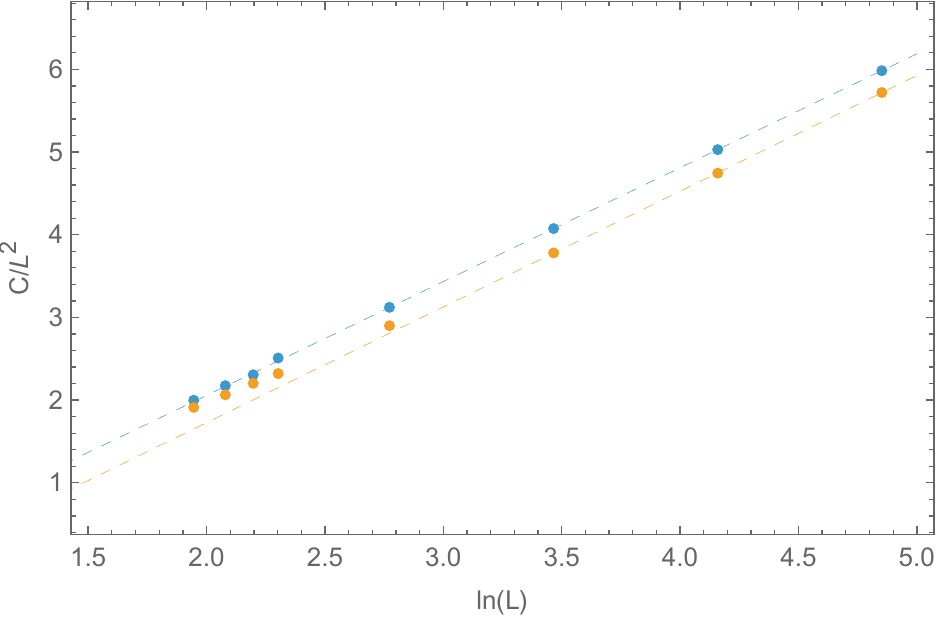}
    \caption{Cost function (\ref{eq:CLA12}) of the generalized Hilbert curves (blue) vs the optimal path (yellow) found by simulated annealing up to $L=128$. The best fit of the data taking the last 4 points ($L=16,32,64,128$) gives $C_{\mathrm{H}}/L^2=1.376\ln L-0.694$ and $C_{\mathrm{O}}/L^2=1.36\ln L-0.90$. The optimal layout conjectured in \cite{mitchison1986optimal} would give $C/L^2=1.422\ln L$. Although the scaling form is almost certainly correct, it seems that both the Hilbert curve and the optimal curve have a smaller cost than hypothesized in \cite{mitchison1986optimal}.}
    \label{fig:cost_HO_vs_L}
\end{figure}

Intuitively, and as discussed in \cite{abedi2025efficient}, a desirable property of the path to optimize the performance of DMRG is that the path distance $\lambda(u,v)$ of a given point $u$ to each of its neighbors on the lattice $v$, is the smallest possible. In fact, these pairs $uv$ connected on the lattice, will typically be entangled with entanglement entropy of $O(1)$. If the lattice points have distance $\lambda\gg 1$ on the given layout, this will require a large bond dimension $\chi$ since, for a chain, it is describing a wave function with {\it volume} entanglement. Therefore one would like to have typically small $\lambda(u,v)$ for $u,v$ close on $G$. Unfortunately, some $\lambda(u,v)$ must necessarily be large of $O(L)$, and one should try to have as few as possible of those, and as many small distances as possible. It turns out that, if we use the previously introduced $\mathrm{LA}_q$, one must choose $q=1/2$ for the optimal paths have exactly this property. 

When computed on a Hamiltonian path, the cost function has a constant contribution of $L^2-1$, since the $L^2$ nearest neighbors both on the lattice and on the path contribute 1 for each bond (independent of $q$). 
Having now settled for $\mathrm{LA}_{1/2}$ as a cost function, we now set to optimize the geometric cost function (dropping the dependence on $G$)
\begin{equation}
    C[\varphi]=\mathrm{LA}_{1/2}(\varphi)-(N-1),
    \label{eq:CLA12}
\end{equation}
where $N$ is the number of of lattice sites, using known methods from statistical physics. In particular we consider minimization of this function in the space of Hamiltonian paths starting and ending at a corner for two different lattices: square and triangular, which are the basis of the models in (\ref{eq:SqLAFM}) and (\ref{eq:TlAFM}) respectively and use the results as a starting point for the DMRG algorithm.

We have investigated this for the square lattice spin-1/2 antiferromagnet in Eq.(\ref{eq:SqLAFM}) looking at various measure of the performance of DMRG for all the Hamiltonian paths in the $6\times 6$ square lattice, selecting 9 representative paths. In Fig.\ref{fig:Correlation_C_DMRG} we present evidence that the geometric cost (\ref{eq:CLA12}) is strongly connected with all measures of performance of DMRG.

In the following section we take this evidence seriously and try to convince the reader that optimizing the geometric cost function (\ref{eq:CLA12}) is a necessary starting point for doing DMRG on higher-dimensional lattices.

\section{Applications to DMRG}

The cost functions (\ref{eq:LAq}) have been introduced in \cite{mitchison1986optimal} and have been considered as proxy of DMRG performance only recently in \cite{bellwood2025fractal}. The main difference between our work and \cite{bellwood2025fractal} is that we will tackle the optimization problem head on, to find the optimal path for any $N$ without recurring to patching curves obtained by the solution of problems with $N'<N$. In this way we can set up and solve to the desired degree of optimality, the problem on a general lattice, a strategy that can be applied to any lattice, and it is not constrained to square lattices.

From the analysis of \cite{mitchison1986optimal} (which is not constrained to paths, but to general layouts), it is clear that the case $q=1/2$ is a special case. In fact, they prove lower bounds for the minimum cost function, which behave like $\mathrm{LA}_{q}\sim L^{1+2q}$ for $1>q>1/2$, $\mathrm{LA}_q\sim L^2$ for $q<1/2$, and $\mathrm{LA}_{1/2}\sim c_1 L^2\ln L$. One can then see that the optimal layout gives a $q-$independent cost for $q<1/2$ and $q=1/2$ is a critical case. The constant $c_1=\frac{\frac{1}{\sqrt{6}}+\frac{1}{\sqrt{3}}}{\log (2)}\simeq 1.422$ obtained for a particular, efficient, construction in \cite{mitchison1986optimal}, is probably not optimal. By taking a good candidate for the minimum, a generalized Hilbert curve ``Gilbert path" (using the code \href{https://github.com/jakubcerveny/gilbert}{at this link}), we find a better scaling. Using these Hilbert curves for $L\times L$ square lattices we find a leading constant $c_1=1.36\div 1.37$. See the supporting data in Figure \ref{fig:cost_HO_vs_L} where we also include the results of our optimization search (more details on this in the following). 

The generalized Hilbert path have shown to be very good choices for the performance of the DMRG algorithm and it is then not surprising that they are very good minima for $\mathrm{LA}_{1/2}$ even though the optimal paths do consistently better.

\begin{figure*}[t]
    \includegraphics[width=1\linewidth]{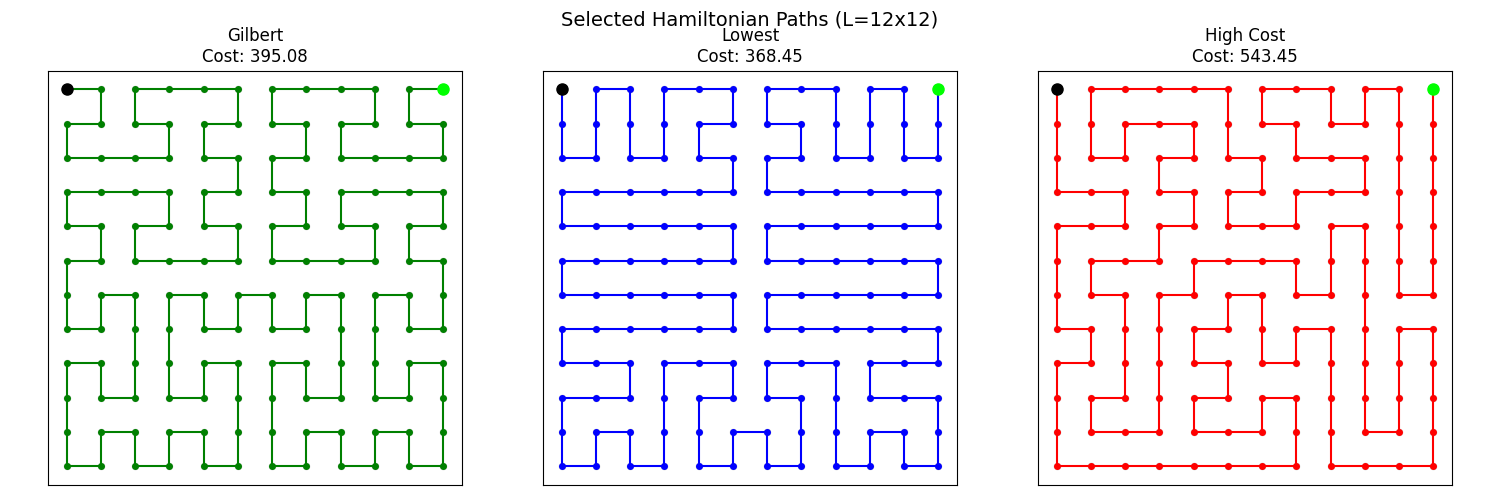}\\
    \includegraphics[width=1\linewidth]{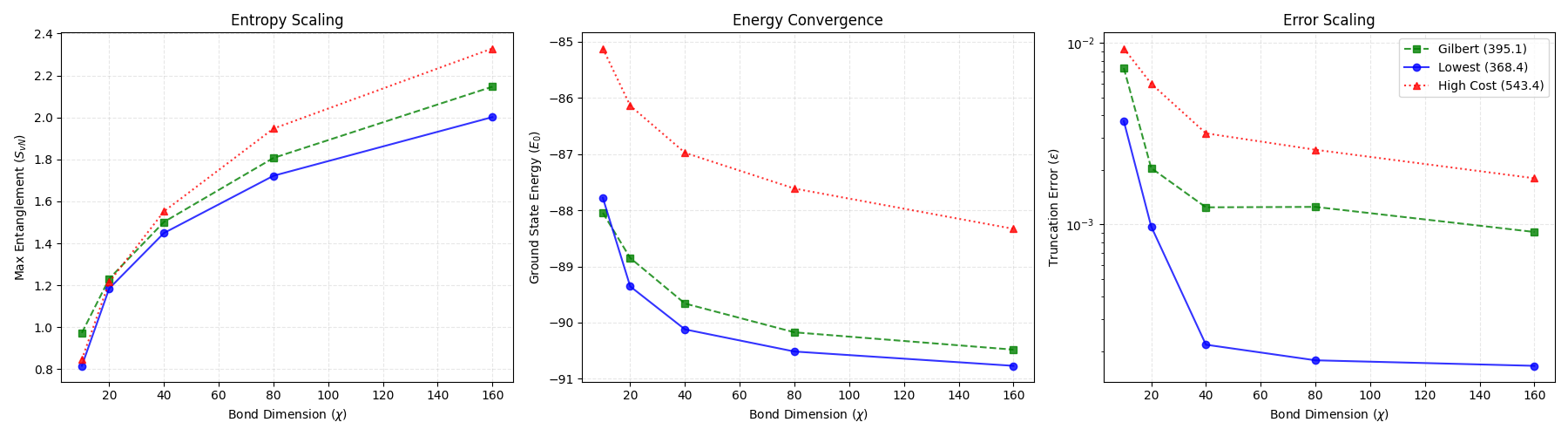}
    \caption{Upper panel: Three paths used for $L=12$.(Left) The Generalized Hilbert (Gilbert) path; (Center) the optimal path; (Right) a high cost path sampled from the Gibbs distribuition with temperature $T=200$.
    Lower panel: Comparison of the performance of DMRG for the ground state of Eq.(\ref{eq:SqLAFM}) as the bond dimension is increased for three paths in $L=12$: Lowest Cost, Generalized Hilbert path and a path sampled from the Gibbs distribution with temperature $T=200$.}
    \label{fig:Three_Paths_12}
\end{figure*}

\subsection{Antiferromagnet on the square lattice}

For the square lattice (see Fig.\ref{fig:Square_Triang}), $N=L^2$ and we set up a simulated annealing algorithm which samples the Boltzmann distribution,
\begin{equation}
    P[\varphi]\propto e^{-\beta C[\varphi]},
\end{equation}
where $\beta$ is gradually increased as the Montecarlo steps $t$ are performed. In the optimization we consider only the Hamiltonian paths which start at one corner and end at another corner of the lattice, with open boundary conditions. It is convenient to choose start and end point at a corner since in this way one cuts only 2 bonds, as opposed to 4 bonds. This is achieved by a generalized Hilbert curve as described before, which already gives a good starting point. Good but not optimal. One starts at an initial larger temperature (mildly dependent on the system size $L$) so starting with $\beta(0)\sim 10$ and increasing to $O(10^3)$ scrambles the original Gilbert path and then cools it down to find a better minimum. 

The \texttt{C++} code, written with help by Gemini Pro starting from a Python code, is available at \href{https://github.com/ascardic/Optimal_Path_DMRG}{this link}. A single restart of the simulated annealing algorithm for $L=64$ takes a second on a MacBook Pro M1 laptop. The path optimization algorithm explores the configuration space of Hamiltonian paths using a local topological update, implemented as the \texttt{mutate\_step} procedure in the class \texttt{Hamiltonian}. This operation iteratively perturbs the site ordering via a two-stage reconnection  process akin to a ``back-bite" move. First, the algorithm performs a \texttt{split}  operation, which identifies a pair of spatially adjacent lattice sites, $u$ and $v$, that are currently non-adjacent in the path sequence; a new link is introduced 
between them, temporarily creating a loop structure. To satisfy the Hamiltonian constraint, a subsequent \texttt{mend} operation identifies and removes a distinct  existing edge, effectively breaking the loop and restoring a single, continuous path that visits every site exactly once with an altered geometry. The end points are left untouched. In Figure \ref{fig:Hilb_vs_SA} we propose several optimal paths for different $L$ with their corresponding Generalized Hilbert paths.

The DMRG algorithm is implemented in \texttt{TenPy}, with a sufficient number of iterations to obtain convergence for the given $\chi$.

We first study the AFM on the square lattice Eq.(\ref{eq:SqLAFM}) and we find encouraging results (see Fig.\ref{fig:Three_Paths_12}). The optimal path is sufficiently different from the Generalized Hilbert path to give at the same time a gain in the energy, maximum entropy, and truncation error for sizes $L\simeq 10$ already. The gain stays, increasing the bond dimension or, equivalently, one can use a smaller bond dimension $\chi$ to achieve better accuracy. In particular, for $L=12$ one observes that practically {\it half} the bond dimension is sufficient to match the accuracy of the Hilbert path. Since the computational cost is proportional to $\chi^3$, this speeds up the DMRG by a large, constant factor $\sim 10$. 

We reiterate that the overhead for finding an optimal path is but a few seconds on laptop, and for some lattices a precompiled list can be done once and for all.

\subsection{Disordered systems on the square lattice and other lattices}

Disordered quantum systems have received a great deal of attention in the last decade, thanks mostly to discovery of Many-Body Localization, which is an emergent integrable phase where unusually slow dynamics is observed (see \cite{sierant2025many} for a recent review). The geometric optimal path can be used for studying disordered systems as well. Eventually one would like to study finite-temperature dynamics, but for the moment, let us consider the problem of finding the ground state. 

\begin{figure*}
    \centering
    \includegraphics[width=0.32\linewidth]{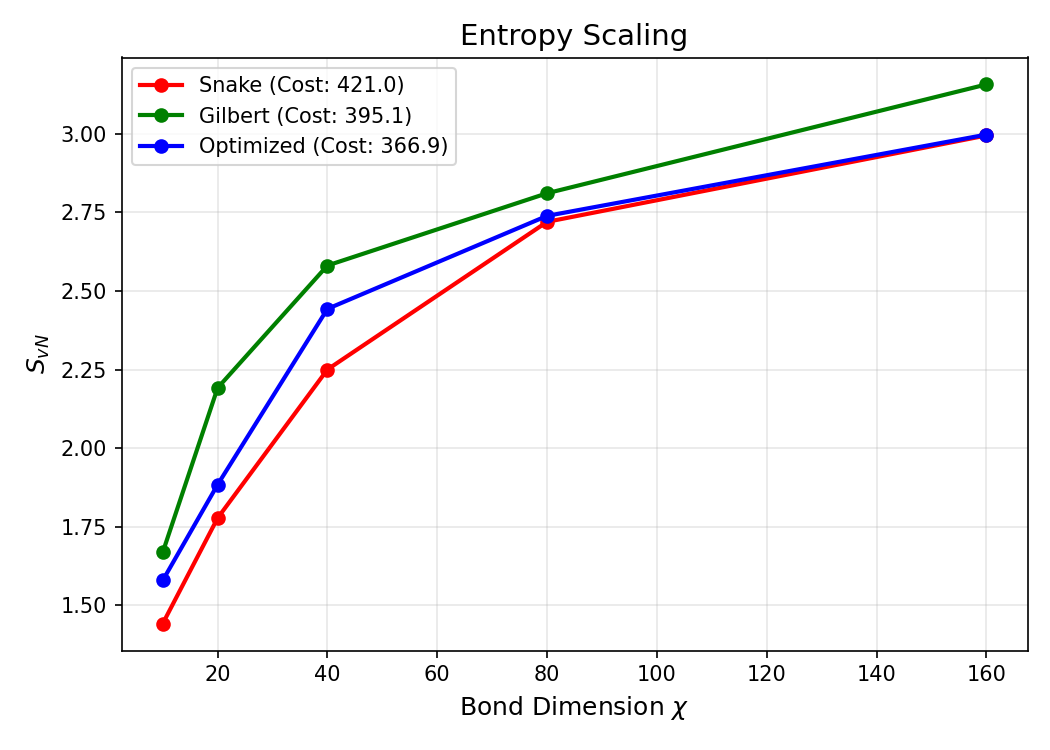}
    \includegraphics[width=0.32\linewidth]{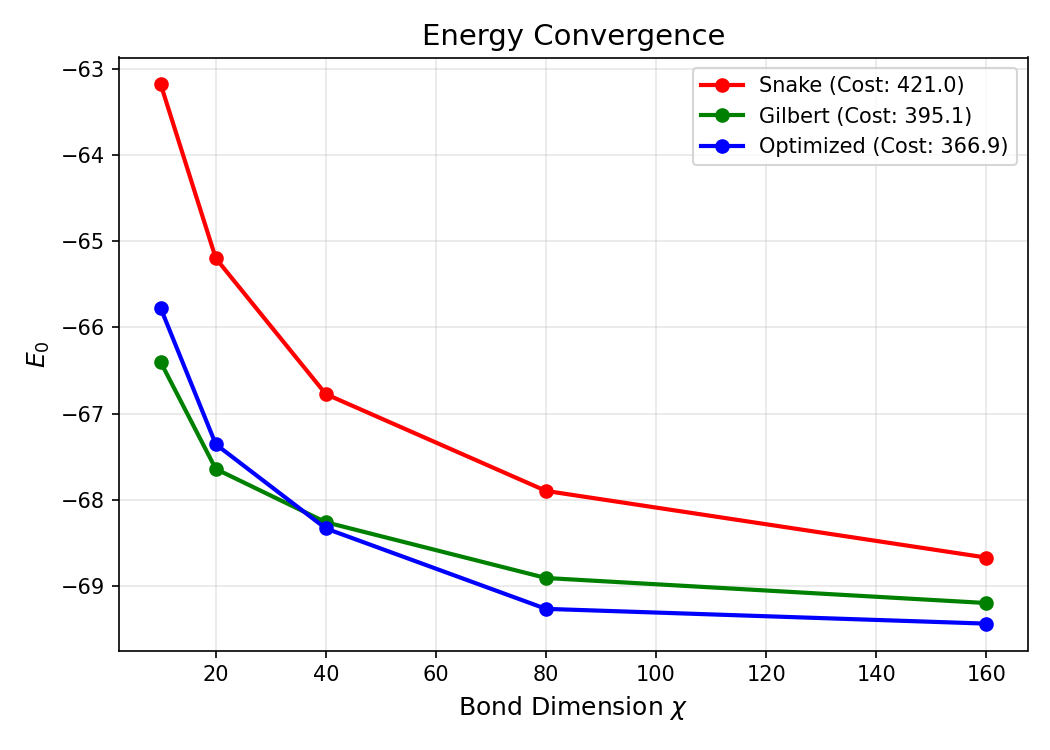}
    \includegraphics[width=0.32\linewidth]{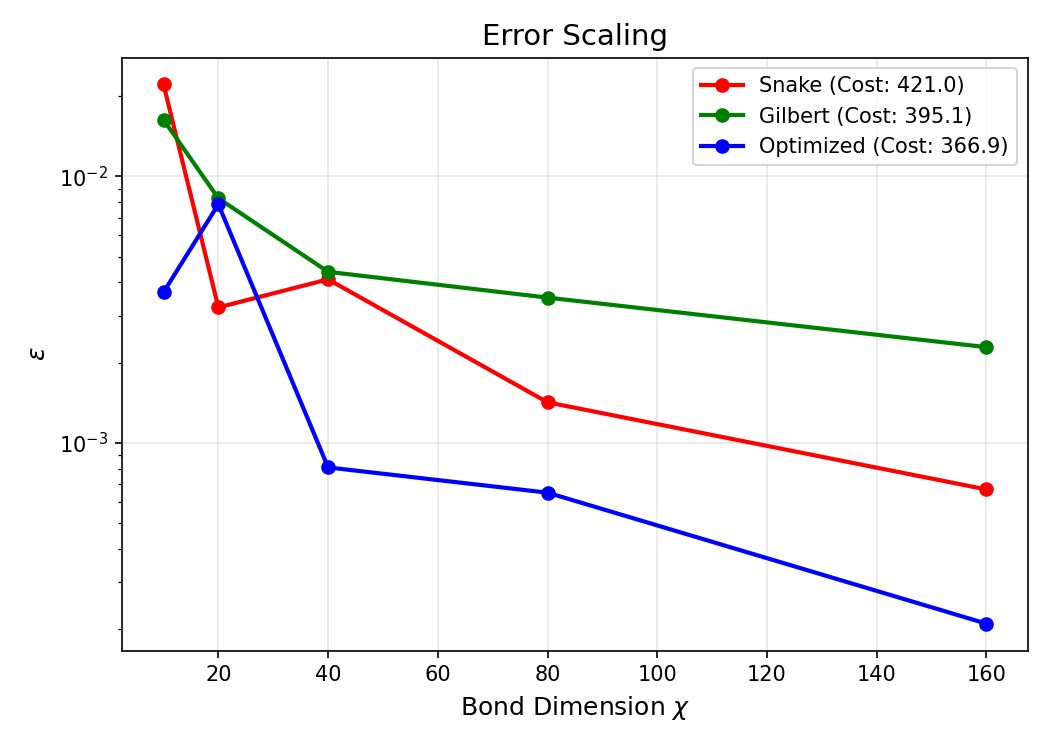}
    \caption{Comparison of the performance of the DMRG to find the ground state of the Heisenberg spin glass in Eq.(\ref{eq:SqL_SG}) using 3 different paths: The traditional snake path, the generalized Hilbert path, and the optimal path.}
    \label{fig:SG_comparison}
\end{figure*}

The performance gain, using the optimal path to find the ground state with DMRG is also significant, which shows that the method has promise to extend beyond the study of clean systems. As a first example, let us consider the spin glass Eq.(\ref{eq:SqL_SG}). The results for a representative, single disorder realization are in Figure \ref{fig:SG_comparison}. The gain in performance is similar to the one on the AFM Heisenberg model in Eq.(\ref{eq:SqLAFM}).

\begin{figure}
    \includegraphics[width=0.95\linewidth]{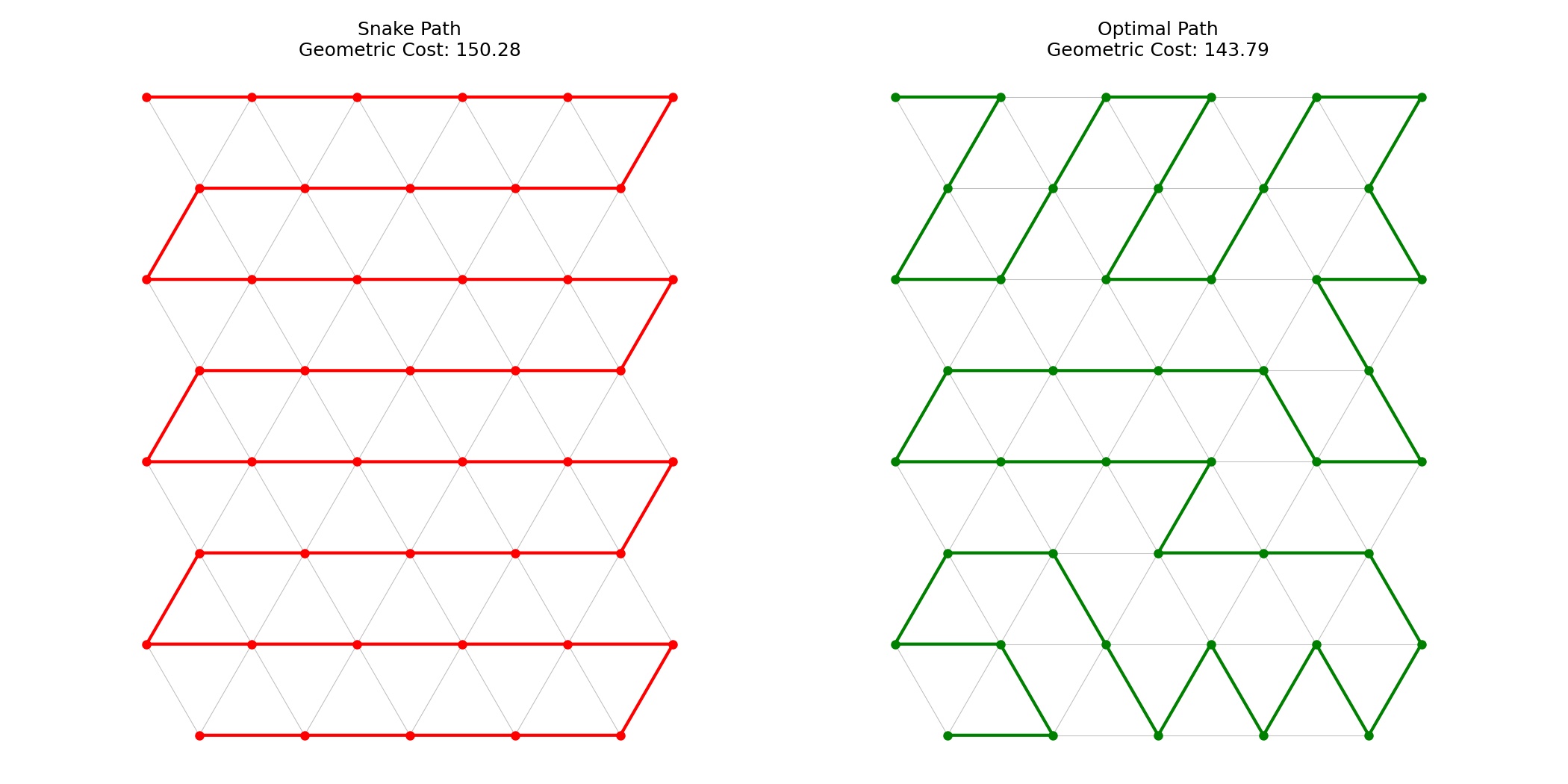}
    \caption{Snake path and best path on the triangular lattice $6\times 8$.}
    \label{fig:Snake_Opt_Triang}
\end{figure}

\begin{figure*}
    \includegraphics[width=0.9\linewidth]{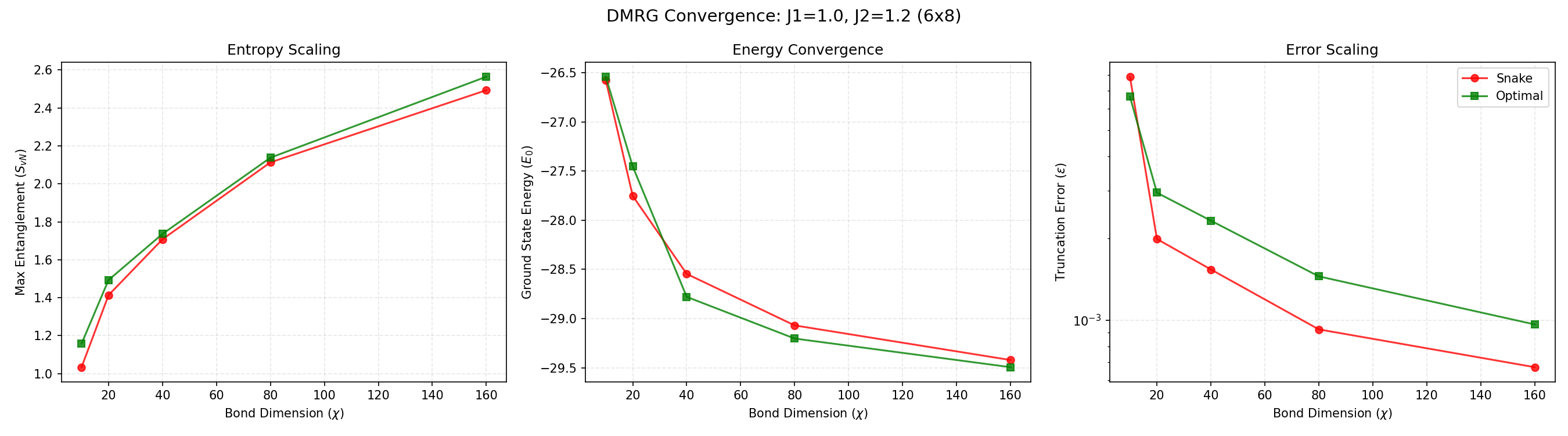}
    \includegraphics[width=0.9\linewidth]{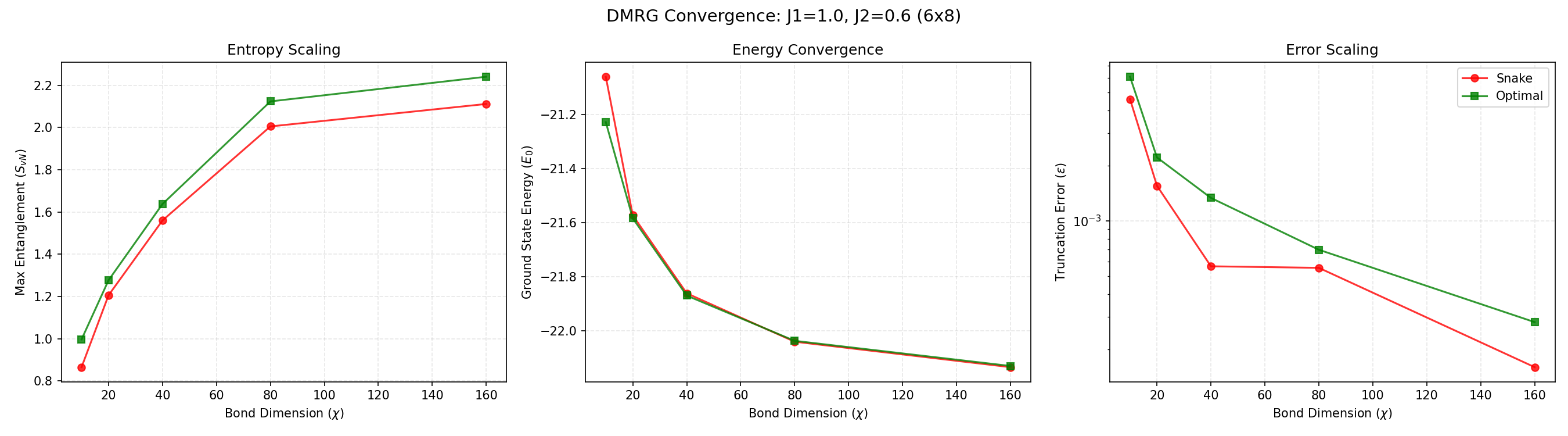}
    \caption{Best path for the triangular lattice and its performance as a layout for the $J_1-J_2$ model. We have $J_1=1$ and from top to bottom $J_2=1.2, 0.6$.}
    \label{fig:Snake_Opt_J1J2}
\end{figure*}

As a second example, we consider the case of non-isotropic lattices like the triangular lattice of the model in Eq.(\ref{eq:TlAFM}). Here we see things becoming more complicated. If one applies the previous strategy, finding the best path for $\mathrm{LA}_{1/2}$ independent of $J_2/J_1$ one does not capture the subtle dependence of the ground state on $J_2/J_1$, and one finds itself in the extreme case in which when $J_2=0$ the best path does not reduce to the best square lattice path, as it should. It is clear then that one can not get a substantial improvement over the snake path. Rather a marginal one is obtained the closer $J_2/J_1$ is to 1. See Figure \ref{fig:Snake_Opt_Triang} for the best path and Figure \ref{fig:Snake_Opt_J1J2} for the results of DMRG run on those paths for a $6\times 6$ section of a triangular lattice. When $J_2/J_1=0.6$ the best and snake path give very similar results.

One must also notice that for the triangular lattice, reducing the temperature quickly in the simulated annealing does not work particularly well, and it is convenient to just generate random paths (running sampling at large temperature) and select the best out of those. The optimization problem on the triangular lattice seems to run into some very rugged landscape and requires further work to be sorted out. We leave this for future work. 

\section{Conclusions}

\begin{figure*}
    \includegraphics[width=0.42\linewidth]{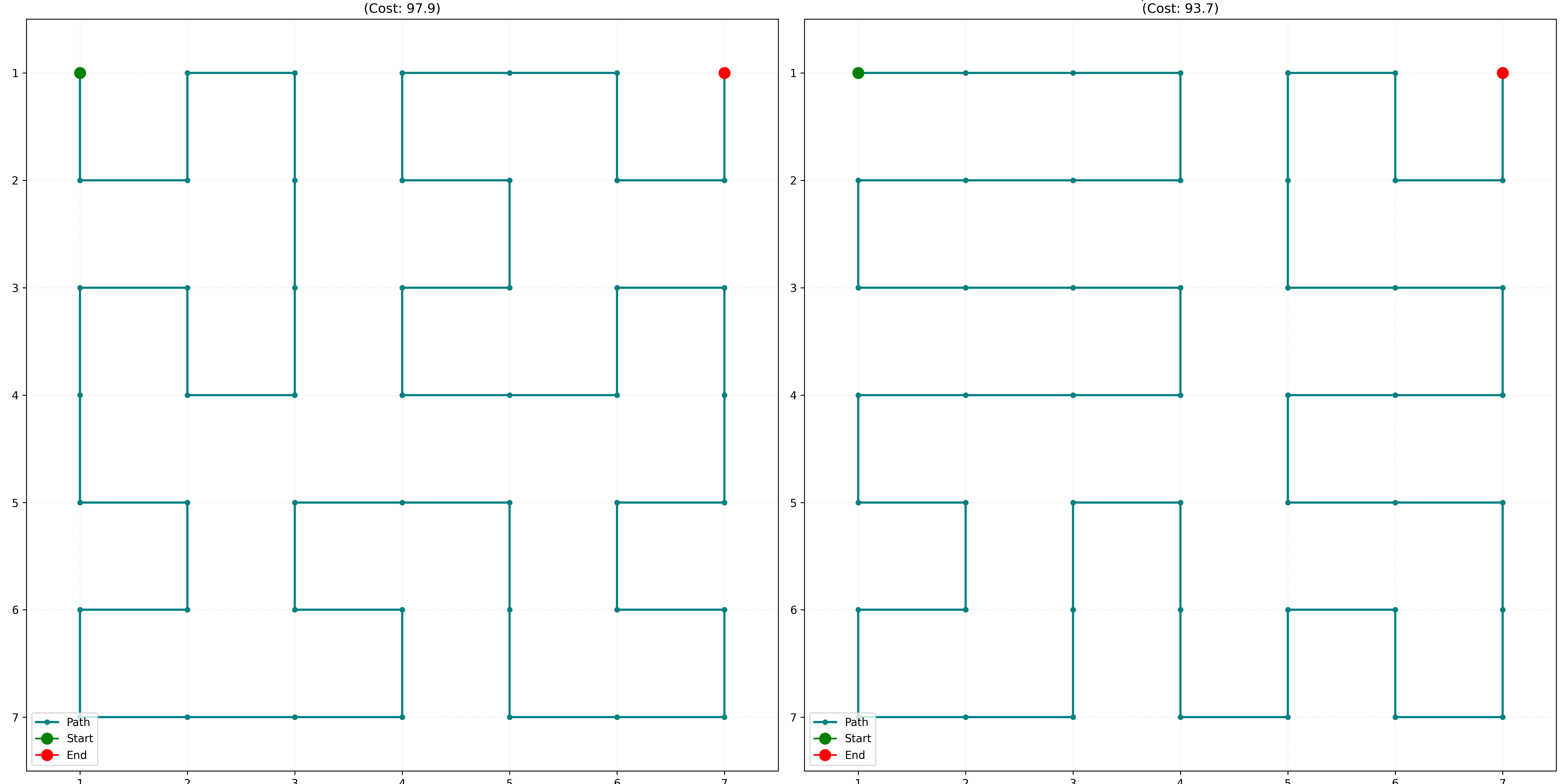}
    \hspace{0.5cm}
    \includegraphics[width=0.42\linewidth]{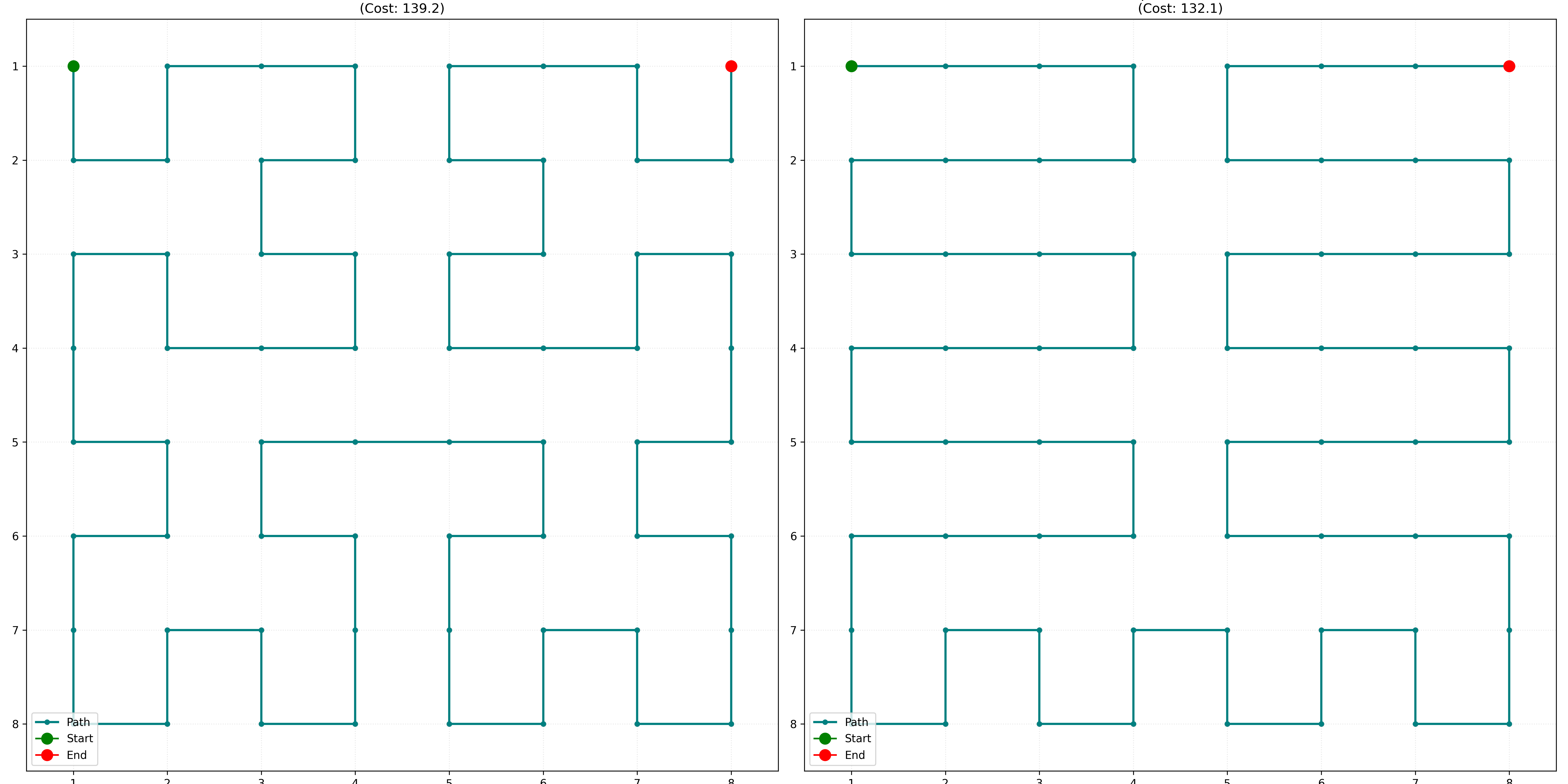}\\
    \vspace{0.5cm}
    \includegraphics[width=0.42\linewidth]{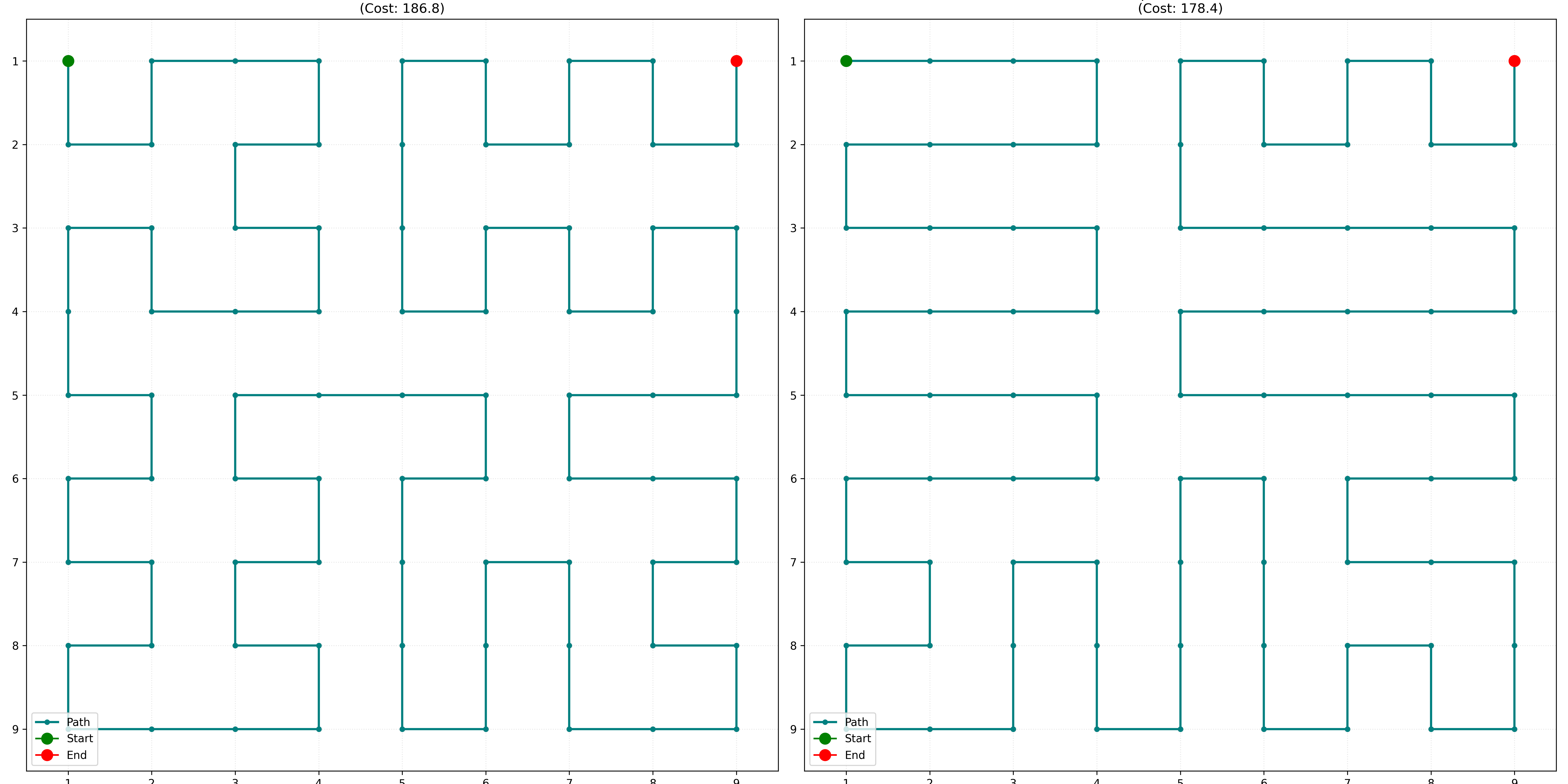}
    \hspace{0.5cm}
    \includegraphics[width=0.42\linewidth]{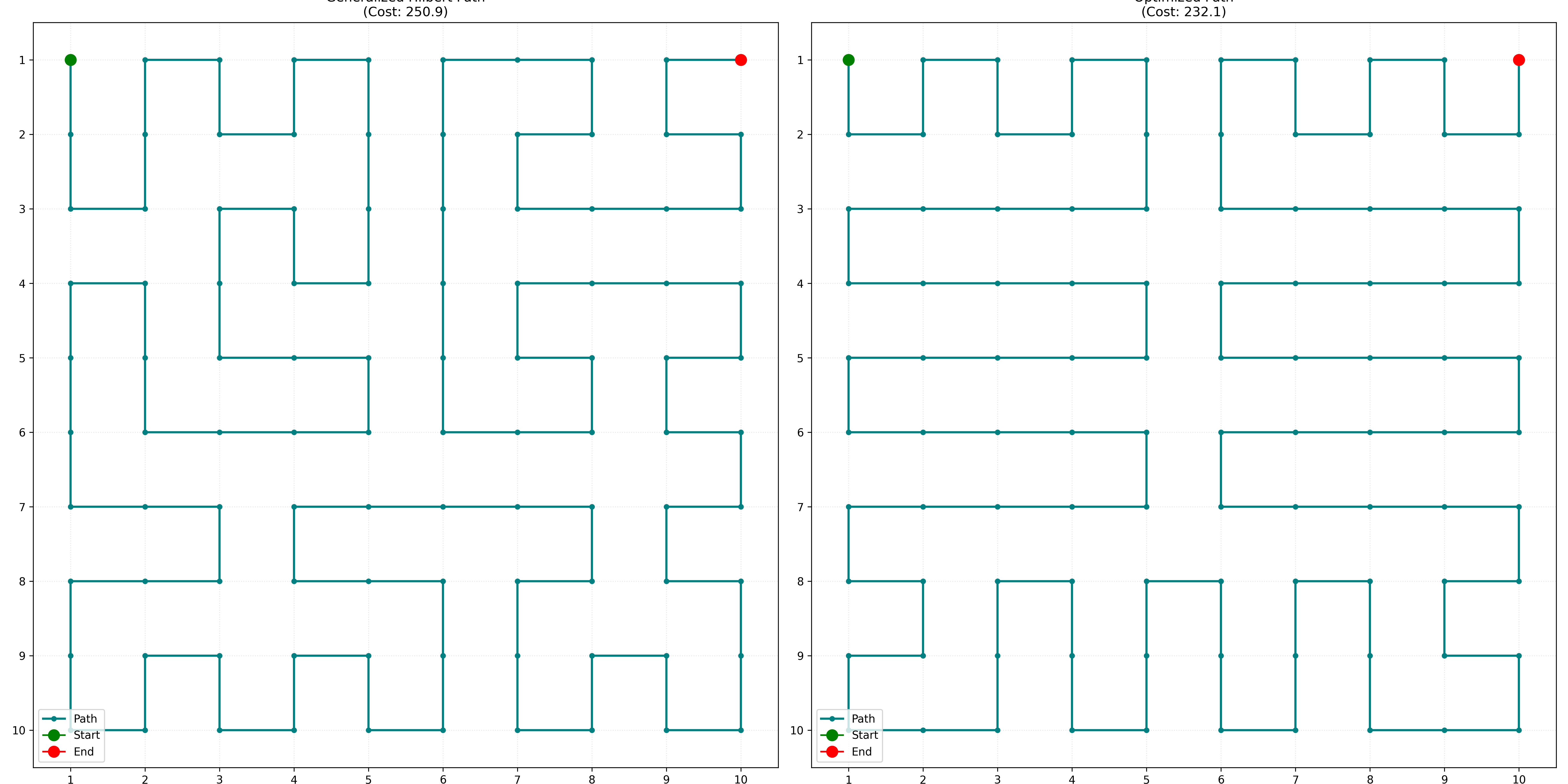}\\
    \vspace{0.5cm}
    \includegraphics[width=0.42\linewidth]{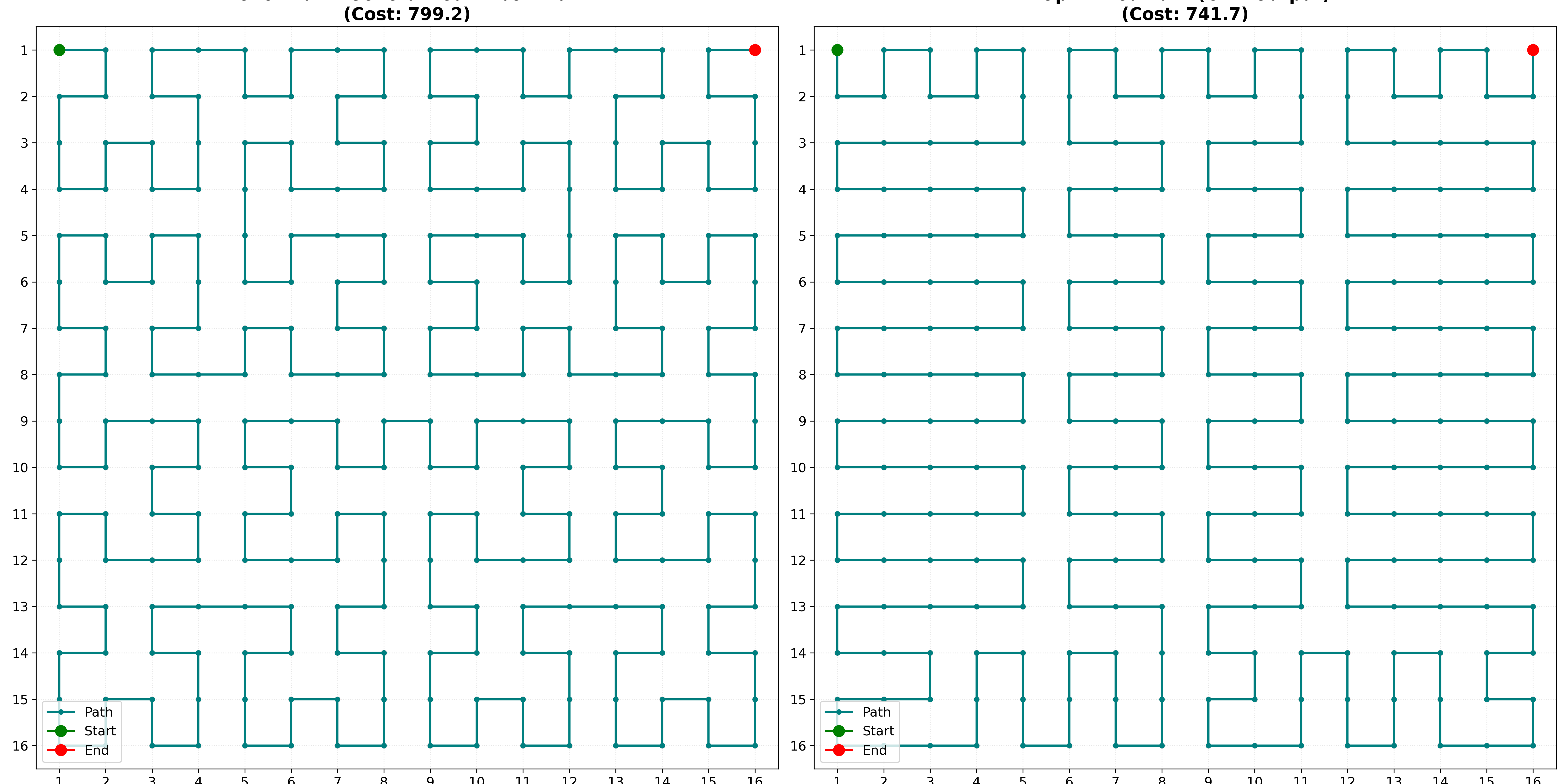}
    \hspace{0.5cm}
    \includegraphics[width=0.42\linewidth]{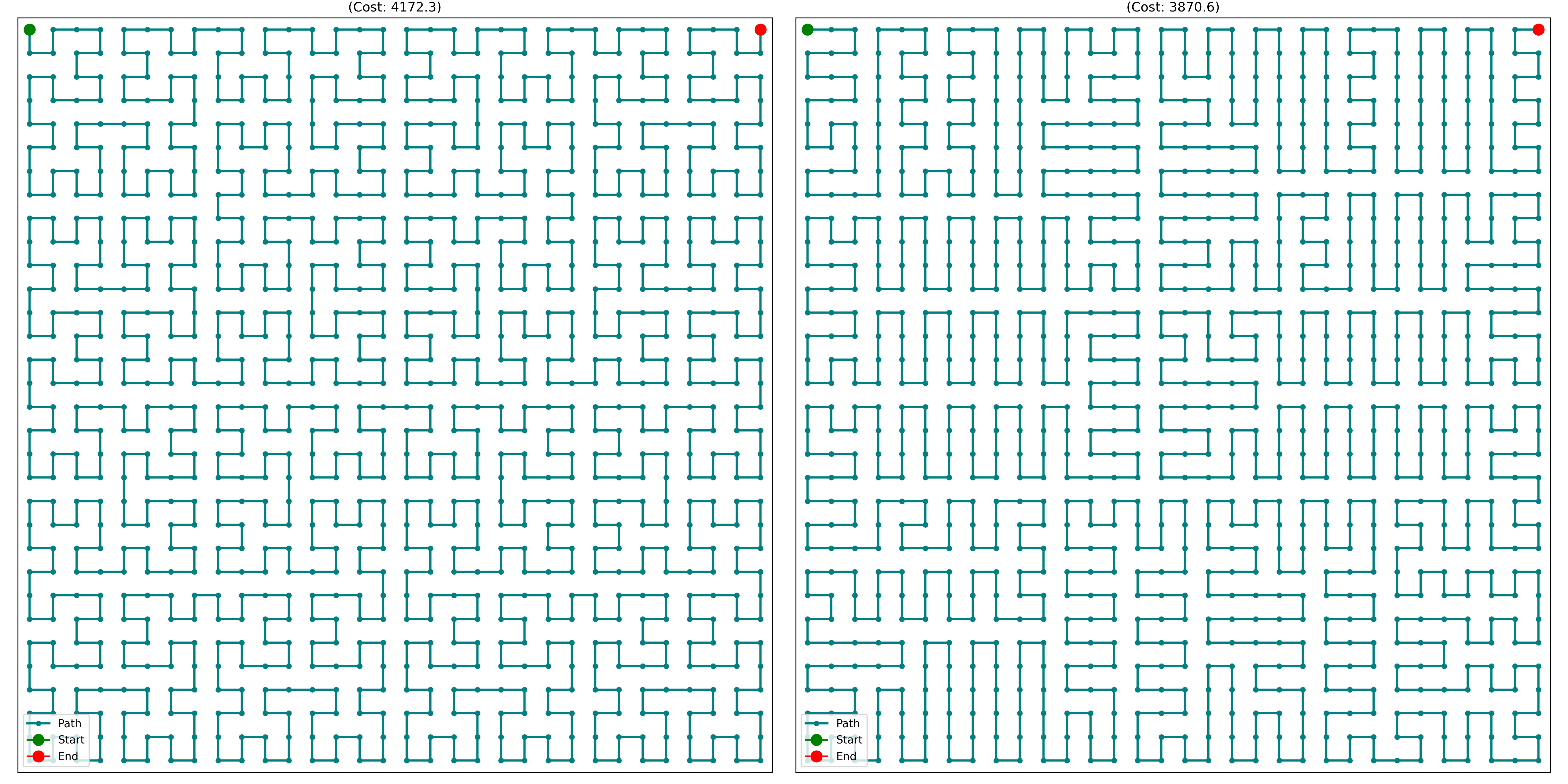}
    \caption{Side by side comparison for the generalized Hilbert path and the optimal path found using simulated annealing with split and mend moves for $L=7,8,9,10,16,32$ (top left to bottom right). We are confident that up to $L\lesssim 20$ the simulated annealing finds the true minimum, while for $L> 20$ a very good minimum is found but possibly not the optimum. The optimization can be extended up to $L=128$ without much computational effort, but the visualization of the path would require more than a page.}
    \label{fig:Hilb_vs_SA}
\end{figure*}

We have investigated the optimal layout to use DMRG/MPS algorithms for two-dimensional lattices. We set it up as an optimization problem of a geometric cost function which is a variation of the Minimum Linear Arrangement problem in graph theory. We set up a simulated annealing algorithm to optimize the layout/path and show that the optimized path improves the DMRG algorithm performance significantly over the default snake path and even over the more reasoned heuristics presented in recent papers, allowing one to sometime achieve the same precision using half the bond dimension of those. We conclude with a list of best paths for various system sizes in Figure \ref{fig:Hilb_vs_SA}. Future directions for the continuation of this work are clearly the investigation of different lattices and the devise of different cost functions that can exploit better the peculiarities of the Hamiltonian under scrutiny. For example, in non-homogeneous Hamiltonians, one should be able to go beyond purely geometric considerations to improve the DMRG performance. In general, we think it is time to explore the optimization of the path choice for DMRG in a more systematic way and we hope the observations provided in this paper will turn useful for the rapidly evolving field of numerical classical simulations.

\section{Acknowledgements}

I am indebted to Marcello Dalmonte, Simone Montangero, and Giuseppe Magnifico for interesting  discussions, comments, and for suggesting some relevant references. The DRMG explorative numerics has been carried over on an Apple Mac Mini M2 using \texttt{TenPy} \cite{tenpy2024}.

\bibliography{bibliography}

\section{Appendix}

We show here the best paths on the $L\times L$ lattice obtained using a simulated annealing code \href{https://github.com/ascardic/Optimal_Path_DMRG}{at this link}. We are confident that the best paths are presented up to $L=20$, while very good but possibly not optimal paths are presented for $L>20$. The code starts from a generalized Hilbert path and then does a simulated annealing search with exponential reduction of the temperature for each step $n=1,...,N$, $T_n=T_0(1-\epsilon)^n$. Typical parameters are: maximum number of steps is $N\sim 10^3\times L^2$, $\epsilon=10^{-5}$, $T_0=10^2$, and the number of restarts is $10^3$. Restarts are parallelized using OpenMP. The parameters, and possibly the annealing schedule, need to be adjusted as a function of $L$. When the same minimum is found a large fraction of restarts one can be reasonably confident one has found the true minimum. This occurs for $L\leq 20$ with the above parameters.

\end{document}